\documentclass[printer]{aa}
\usepackage{graphicx,amssymb}

\def\dof{{\rm{d.o.f.}}}

\begin{document}

\title{
Intrinsic and dust-induced polarization in gamma-ray burst afterglows:
the case of GRB\,021004
\thanks{Based on observations made with ESO telescopes at the Paranal
Observatory under programme Id 70.D-0111, on data from the ESO/ST-ECF
Science Archive Facility and on observations made with the TNG under
programme TAC\,8\_01(47).}}

\author{D. Lazzati\inst{1} \and S. Covino\inst{2} \and S. di Serego
Alighieri\inst{3} \and G. Ghisellini\inst{2} \and J. Vernet\inst{3} \and
E. Le Floc'h\inst{4} \and D. Fugazza\inst{5} \and S. Di Tomaso\inst{5}
\and D. Malesani\inst{6} \and N. Masetti\inst{7} \and E. Pian\inst{8}
\and E. Oliva\inst{5} \and L. Stella\inst{9}}

\titlerunning{Polarization of GRB\,021004}
\authorrunning{Lazzati et al.}

\offprints{D. Lazzati; \email{lazzati@ast.cam.ac.uk}}
\institute{Institute of Astronomy, University of Cambridge, Madingley Road,
CB3 0HA Cambridge, UK.
\and
INAF -- Osservatorio Astronomico di Brera, via E. Bianchi 46, 23807
Merate (LC), Italy.
\and
INAF -- Osservatorio Astrofisico di Arcetri, Largo E. Fermi 5, 50125
Firenze, Italy.
\and
Service d'Astrophysique, C.E. Saclay, 91191 Gif--sur--Yvette Cedex,
France.
\and
INAF -- Telescopio Nazionale Galileo, Roque de los Muchachos, P.O.  Box
5653, 38700 Santa Cruz de la Palma, Spain.
\and
International School for Advanced Studies (SISSA-ISAS), via Beirut 2-4,
34014 Trieste, Italy.
\and
Istituto di Astrofisica Spaziale e Fisica Cosmica, CNR, via Gobetti 101,
40129 Bologna, Italy.
\and
INAF -- Osservatorio Astronomico di Trieste, via Tiepolo 11, 34131
Trieste, Italy.
\and
INAF -- Osservatorio Astronomico di Roma, via Frascati 33, 00040
Monteporzio Catone (Roma), Italy.}

\date{}

\abstract{
Polarization measurements for the optical counterpart to GRB\,021004
are presented and discussed. Our observations were performed with the
TNG and the VLT--UT3 (Melipal) during the first and fourth night after
the gamma-ray burst discovery. We find robust evidence of temporal
evolution of the polarization, which is therefore, at least partially,
intrinsic to the optical transient. We do not find convincing evidence
of wavelength dependence for the intrinsic polarization of the
transient, in agreement with current polarization models for optical
afterglows. We discuss the role of dust, both in our galaxy and in the
host, in modifying the transmitted polarization vector, showing
how a sizable fraction of the observed polarized flux is due to
Galactic selective extinction, while it is not possible to single out
any clear contribution from dust in the host galaxy. We discuss how
our data compare to those obtained by different groups showing that a
two-component model is required to describe the complete dataset. This
is not surprising given the complex lightcurve of GRB\,021004.
 }

\maketitle

\section{Introduction}
\label{sec:int}

It is now well established that gamma-ray burst (GRB) optical
afterglows (OA) can show some degree of linear polarization. To date,
in five cases a positive detection was obtained: GRB\,990510 (Covino
et al. \cite{Co99}; Wijers et al. \cite{WVG99}), GRB\,990712 (Rol et
al. \cite{Rol00}), GRB\,020405 (Bersier et al. \cite{Be02}; Covino et
al. \cite{Co03Apr}; Masetti et al. \cite{Ma03}), GRB\,020813 (Barth et
al. \cite{Ba03}; Covino et al. \cite{CMG02}), and GRB\,030329 (Covino
et al. \cite{Co03c}; Efimov et al. \cite{Ef03}; Magalh\~aes et
al. \cite{MP03}). Usually, the polarized flux is not large, $P\la3\%$,
but in most cases it has been possible to rule out that the observed
polarization was induced\footnote{Here and in the following the only
dust induced polarization that we consider is the one due to the
dichroism of the aligned grains and not the polarization induced by
scattering.} by dust along the line of sight in our own Galaxy
(e.g. Covino et al. \cite{Co99},~\cite{Co03b}).  More recently, it was
also possible to exclude, at least for some cases, a major
contribution to the observed polarization level due to the
interposition of dust in the host galaxy.  In fact, interstellar
polarization is necessarily associated with reddening, and by
modelling the spectral shape of the OA, the total amount of dust
interposed on the line of sight can be constrained, allowing us to put
limits on the dust-induced polarization.

This method is however model dependent, and relies also on the
knowledge of the dust properties in the GRB environment (see
e.g. Lazzati et al. \cite{LCG02}).  Given the present uncertainties,
the possibility that for at least some OA a sizable fraction of the
observed polarization is induced by dust in the GRB environment or in
the host galaxy cannot be excluded yet.  In principle, there are at
least two safe ways to unambiguously detect intrinsic polarization:
the first is to perform multiple observations, looking for temporal
variation of the polarization degree and/or position angle.  Such a
polarization variability would also provide a direct link between the
dynamics of the fireball evolution and the geometry of the emitting
region (Ghisellini \& Lazzati \cite{GL99}, hereafter GL99; Sari
\cite{S99}; Granot et al. \cite{GP02}).  The second is to study the
wavelength dependence of the polarization, possibly extending it to
the infrared, in an attempt to exclude that it follows the ``Serkowski
curve'' typical of Milky Way (MW) interstellar dust polarization
(Serkowski et al. \cite{SMF75}), bearing in mind, however, that the OA
emission could be intrinsically polarized in a wavelength dependent
way.

Recently, some steps in this direction were performed. First, for
GRB\,020813 secure polarization variability was detected, the degree
of polarization decreasing on a day time scale from $P \sim 2\%$
(Barth et al. \cite{Ba03}) to $P \sim 0.8\%$ (Covino et
al. \cite{CMG02}), at a fixed polarization angle. Moreover, again for
GRB\,020813 (Barth et al. \cite{Ba03}), and very recently for
GRB\,030329 (Covino et al. \cite{Co03c}), spectropolarimetry could be
performed; in both cases, small but significant wavelength dependence
was found.

GRB\,021004 was localized on 2002 October 4 at 12:06:14 UT by the
HETE--II satellite (Shirasaki et al. \cite{SGM02}). In the FREGATE
8--40\,keV and in the WXM 2--25\,keV bands the burst had a duration of
about 100 seconds. It thus belonged to the class of long-duration
GRBs.  The optical counterpart was identified less than 10~min after
the burst (Fox \cite{Fo02}) as an $R \sim 15.3$ fading object at the
coordinates $\alpha_{2000} = 00^{\rm h}26^{\rm m}54\fs69$,
$\delta_{2000} = +18\degr55\arcmin41\farcs3$. The early detection of
the OA and its brightness allowed a dense sampling of the light curve
(see e.g. Lazzati et al. \cite{LR02} and references therein), the
identification of several absorption systems, and of a prominent
emission feature in the optical spectra identified as a Ly$\alpha$
emission line from the host galaxy at a redshift $z = 2.328$ (Mirabal
et al. \cite{MH02},
\cite{MH03}; Matheson et al. \cite{MG02}; M\o{}ller et al. \cite{MF02};
Schaefer et al. \cite{SG03}).

In the following we will discuss our three polarimetric observations of
GRB\,021004: one performed in the near infrared (NIR) with the
Italian Telescopio Nazionale Galileo (TNG) at the Canary Islands, and
two in the visible band at the VLT.  Furthermore, we analyzed a publicly
available spectropolarimetric dataset retrieved from the ESO VLT Science
Archive.  We will then compare these measurements, including the
polarimetric observations performed by Rol et al. (\cite{Rol03}), with
theoretical models.

\section{Observations and analysis}
\label{sec:data}

Observations of GRB\,021004 made use of two telescopes. First, the
TNG, equipped with the Near Infrared Camera Spectrometer (NICS) and a
$J$ filter, in the imaging polarimetry mode; second, the ESO VLT--UT3
(Melipal), equipped with the Focal Reducer/low dispersion Spectrometer
(FORS\,1) with a Bessel $V$ filter in the imaging polarimetry mode, and
with the grism 300\,V in the spectropolarimetry mode.

\begin{figure}
\includegraphics[width=\columnwidth]{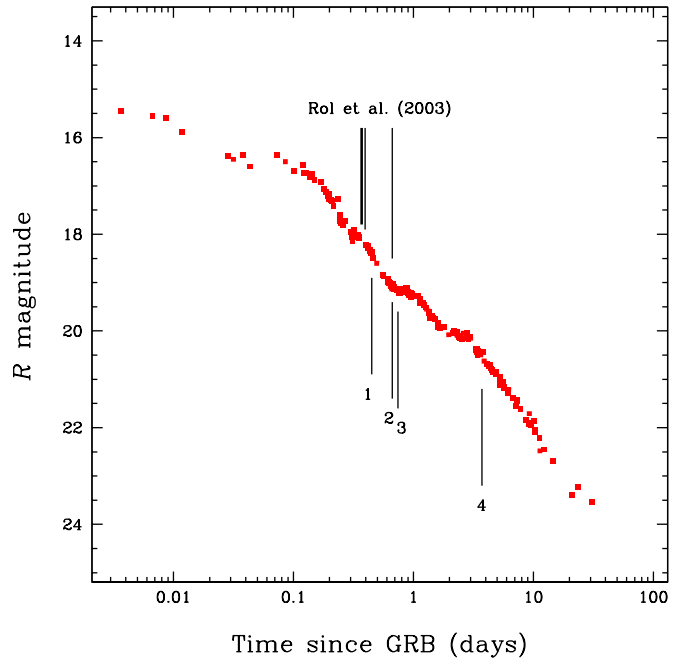}\hfill%
\caption{$R$-band lightcurve of GRB\,021004 with the position of the
polarimetric observations marked.  Data are from Bersier et al
(\cite{Be02}); Fox et al. (\cite{F03}); Holland et al. (\cite{H03});
Pandey et al. (\cite{Pa03}); Uemura et al. (\cite{Ue03}).
\label{fig:polc}}
\end{figure}

\subsection{TNG observation}
The TNG observation (hereafter run~1) started on October 4.915 (9.86
hours after the GRB trigger) and lasted for $\sim 1.8$~hours.  The
optical transient (OT) $J$ magnitude was derived by the acquisition
frames as $J=17.00\pm0.05$ (hereafter 1-$\sigma$ errors are reported).
The observations were performed under mediocre seeing conditions
($1.5\arcsec$) in the large field mode with a scale of
$0.25\arcsec/$pixel.

\begin{figure}[!hb]
\includegraphics[width=\columnwidth]{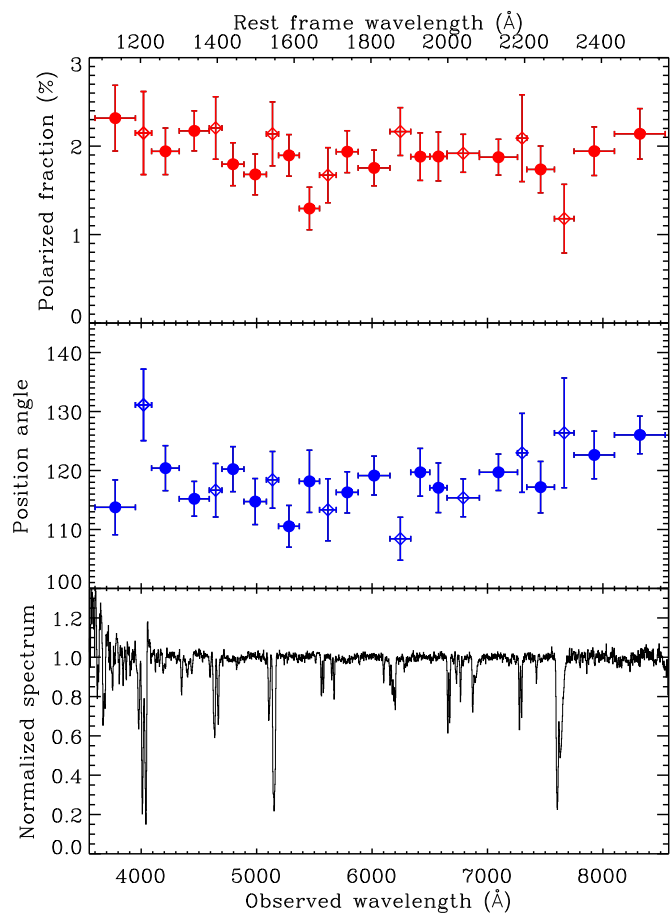}
\caption{Polarization degree, position angle and normalized spectrum
of the public ESO--VLT spectropolarimetric observation (run~3).  The
polarization degree is roughly constant throughout the optical band at
$P \sim 1.9\%$ and the position angle is also constant at $\vartheta
\sim 118\degr$. Observations were performed with the VLT--UT3 (Melipal)
equipped with FORS\,1 in the spectropolarimetry mode with grism
300\,V. Filled symbols correspond to the ``clean'' part of the spectrum,
while empty symbols are relative to absorbed portions of the spectrum,
or the ones contaminated by strong sky emission lines.}
\label{fig:spol}
\end{figure}

\subsection{VLT observations}
We analyzed three sets of polarimetric observations of GRB\,021004 
carried out by the ESO VLT--UT3 (Melipal).

\subsubsection*{First VLT observation}
Our first VLT observation (here called run~2) started on October 5.080
(13.82 hours after the GRB trigger) and lasted for $\sim 1.6$~hours (see
Fig.\,\ref{fig:polc}).  The OT was clearly detected in the acquisition
image with a magnitude $V=19.34\pm0.02$ with respect to the
\mbox{USNO--A2.0} star reported by Fox (\cite{Fo02}), as calibrated by
Henden (\cite{He02a}, \cite{He02b}).

\subsubsection*{Second VLT observation}
From the ESO archive we retrieved a public spectropolarimetric
observation (run~3). It started on October 5.247 (17.83 hours after the
GRB trigger) and lasted for $\sim 2.0$ hours. The spectrum covers
the range from 350~nm to 860~nm and the observations were performed
with the 300\,V grism.

\subsubsection*{Third VLT observation}
Eventually, a last VLT observation (run~4) was performed starting on
October 8.225 (89.3 hours after the GRB trigger), and lasting for
$\sim 2.8$~hours.  The OT magnitude was $V=20.89\pm0.03$ with respect
to the same \mbox{USNO--A2.0} star.

All VLT observations were performed under
good/excellent seeing conditions ($0.5\arcsec-0.9\arcsec$) in standard
resolution mode with a scale of $0.2\arcsec$/pixel.

Polarimetric standard stars were also observed. One polarized,
\mbox{BD-125133}, in order to fix the offset between the polarization
and the instrumental angles, and three non-polarized, \mbox{WD
0310--688}, \mbox{WD 1615--154}, and \mbox{BD+284211}, to estimate the
degree of spurious polarization possibly introduced by the
instruments. In addition, we have also analyzed the ESO archive
spectropolarimentric data for the NGC\,2024 NIR1 polarization standard 
star, obtaining in all cases a good Serkowski curve completely consistent 
with the available
data\footnote{\texttt{http://www.eso.org/instruments/fors1/pola.html}}.

\subsection{Data reduction and analysis}
The data reduction was carried out with the {\tt Eclipse}
package (version 4.3.1; Devillard \cite{D97}). After bias subtraction,
non-uniformities were corrected using flat-fields obtained without
the Wollaston prism. For the IR data sky flat-fields were applied.
The flux of each point source in the field of view was
derived by means of aperture photometry by the Graphical Astronomy and
Image Analysis ({\tt GAIA}) tools%
\footnote{\texttt{http://star-www.dur.ac.uk/$\sim$pdraper/gaia/gaia.html}}
(version 2.6-6).

The general procedure followed for the analysis of imaging polarimetry
observations is indeed extensively discussed in Covino et al. (\cite{Co99},
\cite{CLM02}, \cite{Co03Apr}) and di Serego Alighieri (\cite{dS97}),
while details about the NICS polarimetric capabilities are discussed
by Oliva (\cite{O97}).

The VLT Spectra were extracted with the \texttt{apall} tool included in the
IRAF package (version 2.11), which extracts the spectrum in a
fixed-width window, allowing for a polynomial evolution of the
centroid with wavelength. Suitable IDL routines were also developed to
perform an independent extraction and check for any possible bias. In
this case the extraction of the 1D spectrum from the 2D frame was
performed by fitting a bell-shaped function to all the vertical
stripes of the 2D spectrum, allowing for a first order variation of
the background and for a 4$^{\rm{th}}$ degree polynomial variation of
the flux centroid and width of the fitting function. The result of
this ``fitted'' extraction were not entirely consistent with those of
a more standard sliding window, as discussed below.

The results of the spectropolarimetric observations are shown in
Fig.\,\ref{fig:spol}. To all our data we have applied a correction for
the wavelength dependence of the half-wave plate axis as measured by
the FORS team%
\footnote{\texttt{http://www.eso.org/instruments/fors1/}}.
The plot is based on the fitting extraction (hereinafter F) which, as
anticipated, is not fully consistent with the more standard window
extraction (hereinafter W). The F extraction shown in the figure yields
a polarization which is consistent with being wavelength independent.  A
simultaneous constant fit to the polarization degree and angle returns
$P = (1.88\pm0.05)\%$ and $\vartheta = (118\pm1)^\circ$ with
$\chi^2/\dof=45.8/44$ (null probability $\sim60\%$). The bin size has
been made large enough to make the polarization bias (Wardle \& Kronberg
\cite{WK74}) unimportant.  Completely consistent results are obtained
with the W extraction if a small window (6 pixels width) is adopted,
while a wider window (20 pixels) yields different results. The
polarization obtained using the wide W extraction is wavelength
dependent, the polarization decreasing from $\sim 2\%$ at 600~nm to
$\sim1\%$ at 800~nm. The resulting polarization also presents a
significant dip at $P\sim1\%$ around 550~nm. In principle, both methods
are prone to inadequacies in extracting the data, and it is not possible
to decide {\it a priori} which is the best one. The fit and small W
extraction are in fact sensitive to non centrally symmetric 2D spectra.
The large window, instead, can yield an unbiased extraction of
non-symmetric 2D spectra, but is subject to a varying background and to
contribution from nearby sources. In an ``Ockham razor'' approach, we
decided to show the result which requires less parameters to be
explained, namely the constant one, warning however the reader that
according to the large window extraction, the polarization may be
wavelength dependent, being smaller in the red part of the spectrum.

From an independent analysis of the same data, Wang et
al. (\cite{W03}) report a marginal evidence of increasing polarization
at $\lambda\la400$~nm, across the rest-frame Ly$\alpha$ absorption
features. They interpret this as a consequence of absorption of the
fireball emission by nearby high velocity clumps of HI atoms, with a
covering factor smaller than, but comparable to, unity. As can be seen
in the upper panel of Fig.~\ref{fig:spol}, our binned data do not
support this claim, and the same result is obtained with a binning of
$\sim2$~nm, comparable to the one adopted by Wang et al. (\cite{W03}).
However we find that using smaller wavelength bins (e.g. 0.52~nm) the
polarization, not corrected for the bias (Wardle \& Kronberg
\cite{WK74}), steadily increases from 400 nm down to the shortest
available wavelength as expected for a low sensitivity region at the
blue edge of the spectral range. Therefore this UV rise of the
polarization looks consistent with a bias effect, although the
possibility that some of it might be real cannot be completely
excluded.

We stress that the difference between the various spectrum extraction
methods discussed here are at the $\sim 1\%$ level.  While being small,
if not negligible, for a standard spectral analysis, these differences
become significant when weak polarization levels are investigated.

\begin{table*}[t]
\caption{The normalized polarization Stokes parameters not corrected for the
interstellar polarization. Observations were performed with the TNG
(run~1) and with the VLT--UT3 (runs~2 and 4). VLT--UT3
spectropolarimetry (run~3) was performed with grism 300\,V. The
reported results for the spectropolarimetric observation of run~3 were
obtained by integrating the spectrum over the $V$-band. Uncertainties
are at 1-$\sigma$ and the upper limit is a 95$\%$ confidence level.}
\label{tb:pol}
\centering\begin{tabular}{lcccccccc}\hline
\bf Run &\bf Tel. & \bf Filter &\bf UT (10/02)&\bf Magnitude  &\bf $Q$            &\bf $U$   &\bf $P$ ($\%$)&\bf $\vartheta$ ($\degr$)\\ \hline
1       &TNG & $J$        &4.953  &$17.00\pm0.05$ &$-0.0160\pm0.0130$ &$-0.0222\pm0.0130$ &$<5$          &---        \\
2       &VLT & $V$        &5.172  &$19.34\pm0.02$ &$-0.0083\pm0.0009$ &$-0.0094\pm0.0010$ &$1.26\pm0.10$ &$114\pm2$  \\
3       &VLT & $V$        &5.247  &---            &$-0.0075\pm0.001$  &$-0.016\pm0.001$   &$1.74\pm0.20$ &$122\pm2$  \\
4       &VLT & $V$        &8.225  &$20.89\pm0.03$ &$-0.0067\pm0.002$  &$+0.0002\pm0.002$  &$0.67\pm0.23$ &$89\pm10$  \\
\hline
\end{tabular}
\end{table*}

\subsection{Other observations}

Rol et al. (\cite{Rol03}) obtained three independent measurements of the
optical polarization of the afterglow of GRB\,021004, three with
the Nordic Optical Telescope, in the $R$-band, and one with the VLT, in
the $V$-band. They obtain $P = (1.17\pm0.46)\%$ with position angle
$\vartheta = (184.2\pm11.4)\degr$ at $\Delta{}t = 0.366$~d after the
GRB, $P = (1.73\pm0.51)\%$ with position angle $\vartheta =
(166.4\pm8.1)\degr$ at $\Delta{}t = 0.376$~d, $P < 1\%$ at $\Delta{}t
= 0.396$~d and $P = (1.29\pm0.13)\%$ with position angle $\vartheta =
(121.8\pm2.8)\degr$ at $\Delta{}t = 0.666$~d.

\section{Results and modelling}
\label{sec:model}

\subsection{Imaging polarimetry}

The IR polarimetric observation (run~1) provided $Q$ and $U$ Stokes
parameters compatible with those derived for run~2 (see below and
Tab.\,\ref{tb:pol}), even though the larger errors only allowed us to
derive a $P<5\%$ upper limit ($95\%$ confidence level).  This limit is
not particularly stringent compared to those derived in the optical
for other afterglows (e.g. Hjorth et al. \cite{Hj99}; Covino et
al. \cite{CLM02}; Bj\"ornsson et al. \cite{Bj02}; Covino et
al. \cite{Co03d}).  However, it is the first IR polarimetric
observation that provides a useful constraint, since the previous
upper limits were rather loose ($\sim 30\%$; Klose et
al. \cite{KSF01}).

\begin{figure}
\includegraphics[width=\columnwidth,keepaspectratio]{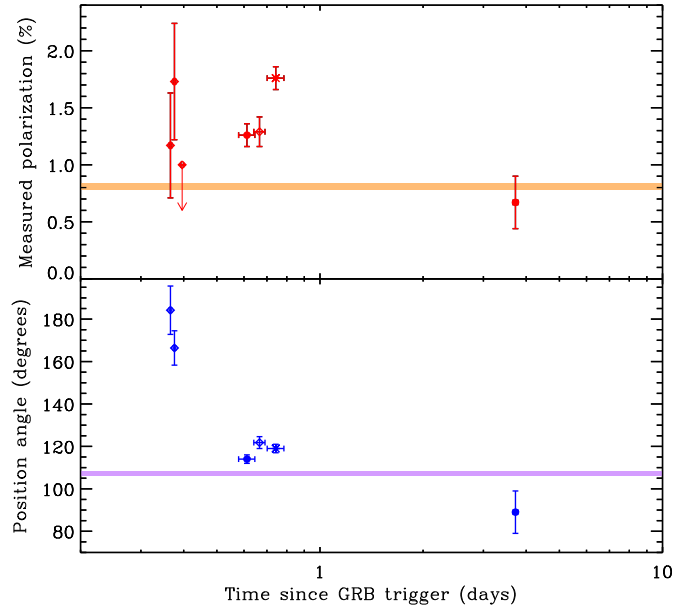}
\caption{$V$-band time-resolved polarimetry of GRB\,021004.
Filled circles show the polarization degree (upper panel) and position
angle (lower panel) for our VLT runs (run 2 and run 4) vs. the time
elapsed since the GRB trigger. The diamonds refer to the Rol et
al. (\cite{Rol03}) measurements, while the asterisk is obtained by
filtering with a Bessel $V$ transmission profile the
spectropolarimetric observation (run 3). The shaded bands show the MW
interstellar polarization for field stars. Their vertical width
corresponds to the 1-$\sigma$ uncertainty.
\label{fig:ipol}}
\end{figure}

The results of VLT $V$-band imaging polarimetry, as well as a
synthetic $V$-band measurement derived from the spectropolarimetric
measurement, are plotted in Fig.~\ref{fig:ipol} as a function of the
time elapsed since the GRB trigger.  We also report the value of
polarization induced by the MW selective extinction, as derived by
averaging the $Q$ and $U$ parameters for several bright stars in the
field.  We obtain $P_{\rm{MW}}=(0.81\pm0.03)\%$ and
$\vartheta_{\rm{MW}}=(107\pm1)\degr$. Such a polarization is slightly
larger than the maximum interstellar (ISM) polarization according to
the empirical law $P_{\rm ISM}\la0.09\,E_{B-V}$ (Serkowski et
al. \cite{SMF75}), given the reddening $E_{B-V}\sim0.06$ estimated for
the GRB\,021004 field (Schlegel et al. \cite{SFD98}).  We note however
that deviations from the above law are observed especially along
low-reddening lines of sight (see Fig.~9 of Serkowski et
al. \cite{SMF75}).  An important conclusion that can be drawn from
Fig.~\ref{fig:ipol} is that neither the polarized fraction nor the
position angle of the OT were constant during the evolution of the
afterglow (Rol et al. \cite{Rol03}).  We conclude that the relative
contribution of the polarizing components, namely the MW ISM, the GRB
host ISM, and the OT itself, did vary on short time scales. Since the
ISM polarization cannot vary on short (day) time scales, we conclude
that the OT was intrinsically polarized.  The rotation of the
polarization angle between the first measurements of Rol et
al. (\cite{Rol03}) and our late-time datum is consistent with
$90\degr$, a quantity predicted in the more commonly accepted models
for the polarizations of GRB jets (GL99; Sari~\cite{S99}). However, in
contrast with the prediction, the transition between the two angles is
not sharp, but rather smooth. This suggest the presence of an
additional polarizing component, as we will discuss below.

\subsection{Spectropolarimetry}

We have shown in Fig.~\ref{fig:spol} the result of our reduction of
the spectropolarimetric observation publicly available at
ESO. Spectropolarimetric observations of OTs are important since they
allow us, in principle, to single out which component is contributing
more to the observed polarization: the OT itself, the host ISM, or the
MW ISM. This is possible since the three components have different
wavelength dependencies.

The OT polarization is supposed to be wavelength independent, at least
in the limited spectral range investigated here. ISM polarization is
instead wavelength dependent. It follows a ``Serkowski law'':
\begin{equation}
P(\lambda)\, =\,
P_{\rm max}\, \exp \left[ -K\,\ln^2
\left( {\lambda_{\rm max} \over \lambda}\right) \right],
\end{equation}
where $P_{\rm max}$ is the maximum induced polarization, obtained at
$\lambda_{\rm max}$. Experimentally,
$0.34~\mu\mathrm{m}\la\lambda_{\rm max}\la 0.9\,\mu$m, and this
parameter is considered to be a measure of the size of the polarizing
grains: the larger $\lambda_{\rm max}$, the larger the grains. For the
numerical value of the coefficient $K$, we follow the more recent
study of Martin et al. (\cite{MCW99}):
\begin{equation}
K\, =\, \left\{
\begin{array}{ll}
1.66 \, \lambda_{\rm{max}} & \lambda \ge \lambda_{\rm{max}}; \\
-0.59+2.56\, \lambda_{\rm{max}} & \lambda < \lambda_{\rm{max}}.
\end{array} \right.
\end{equation}

\begin{table}
\caption{Results of the simple modelling of the spectropolarimetric 
result.  $^\dag$ The constant polarization value is given for the OT
model.  $^\ddag$ The value $\lambda_{\rm max}=0.34$~$\mu$m has been
set as a lower limit to the parameter (see text).
\label{tab:spol}}
\begin{tabular}{|l||llll|}
\hline
Model     &$P_{\rm max}(\%)$ &$\vartheta$ ($^\circ$) &$\lambda_{\rm max}$ ($\mu$m) &$\chi^2/\dof$ \\
\hline
OT$^\dag$ & $1.88\pm0.05$    & $118\pm1$ & ---              & $45.8/44$ \\
MW ISM    & $1.95\pm0.1$     & $118\pm1$ & $0.52\pm0.05$    & $50.5/43$ \\
Host ISM  & $2.1\pm0.15$     & $118\pm1$ & $0.34^\ddag$     & $57.2/43$ \\
\hline
\end{tabular}
\end{table}

To model the polarization results shown in Fig.~\ref{fig:spol}, we have
assumed first that the entire polarization is due to only one of the
three possible components (i.e. OT, MW, host), taking into account the
wavelength dependence of the ISM polarization as detailed above, and
calculating the effect of the host-ISM in the host rest
frame\footnote{The application of the Serkowski law to high-redshift
systems is not directly supported by observations, given the intrinsic
difficulty in performing spectropolarimetric studies of distant
objects. It is however at least partly justified by the observation of
interstellar polarization following a Serkowski curve in SN\,1986G in
Centaurus~A, at $z = 0.00183$ (Hough et al. \cite{H87}).}.  As reported
in Tab.~\ref{tab:spol}, the quality of the data allows us to exclude a
dominant role of the host ISM in producing the observed polarization,
but it is not possible to identify a dominant component between MW-ISM
and intrinsic.  The first two fits are both acceptable ($\chi^2/\dof <
1.2$), with a slight preference for the OT model, while the host-ISM fit
yields a significantly worst $\chi^2$. Performing the fit only on the
high quality dataset (filled symbols in Fig.~\ref{fig:spol}) yields
statistically indistinguishable results. However, since the field stars
do show a moderate degree of polarization, we know that the observed
polarization cannot be entirely due to the OT itself. Furthermore, at
all frequencies the polarization position angle is very similar to that
of the field stars in the imaging polarimetry (see above).

\begin{figure*}
\includegraphics[width=\columnwidth]{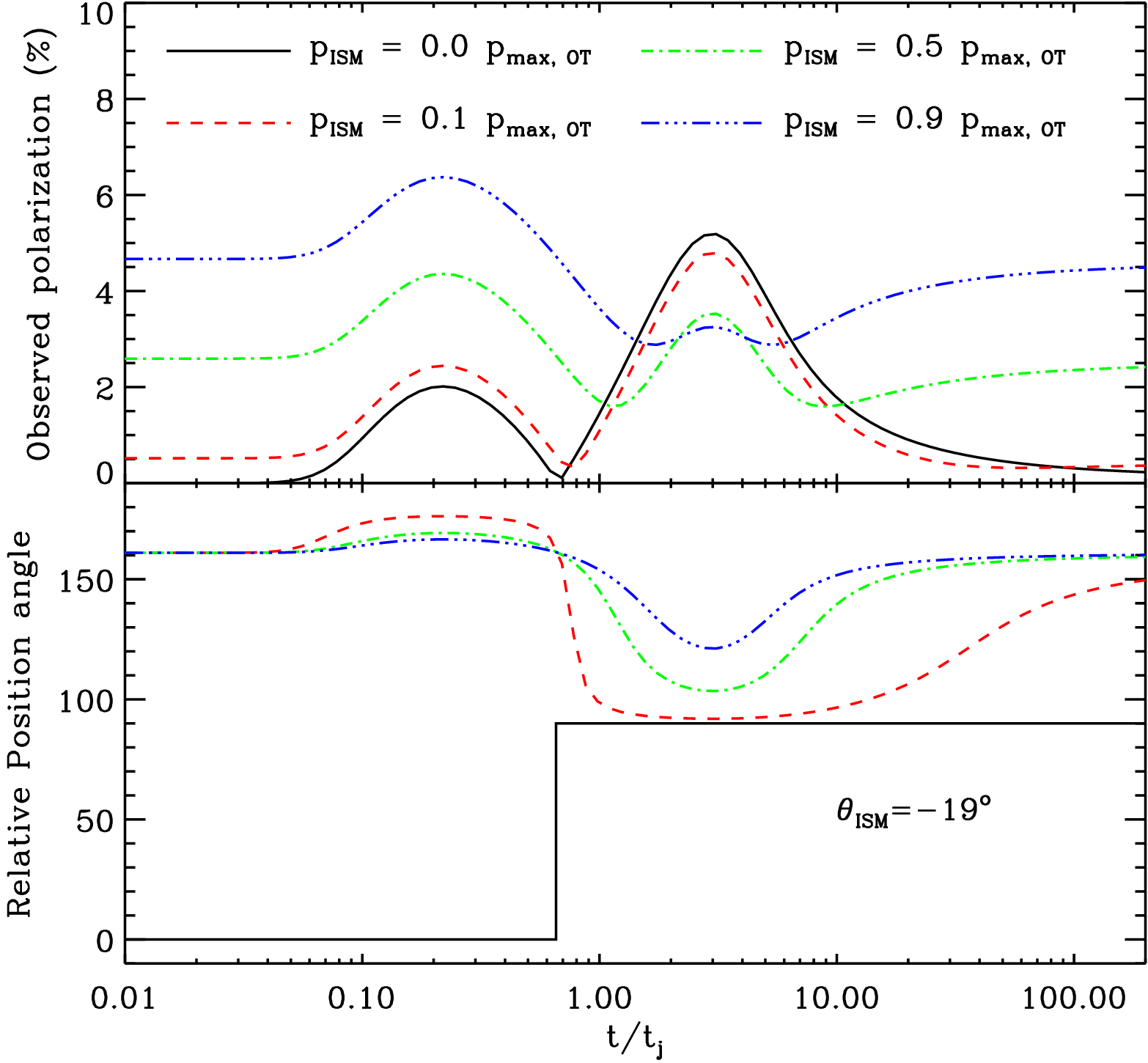}\hfill%
\includegraphics[width=\columnwidth]{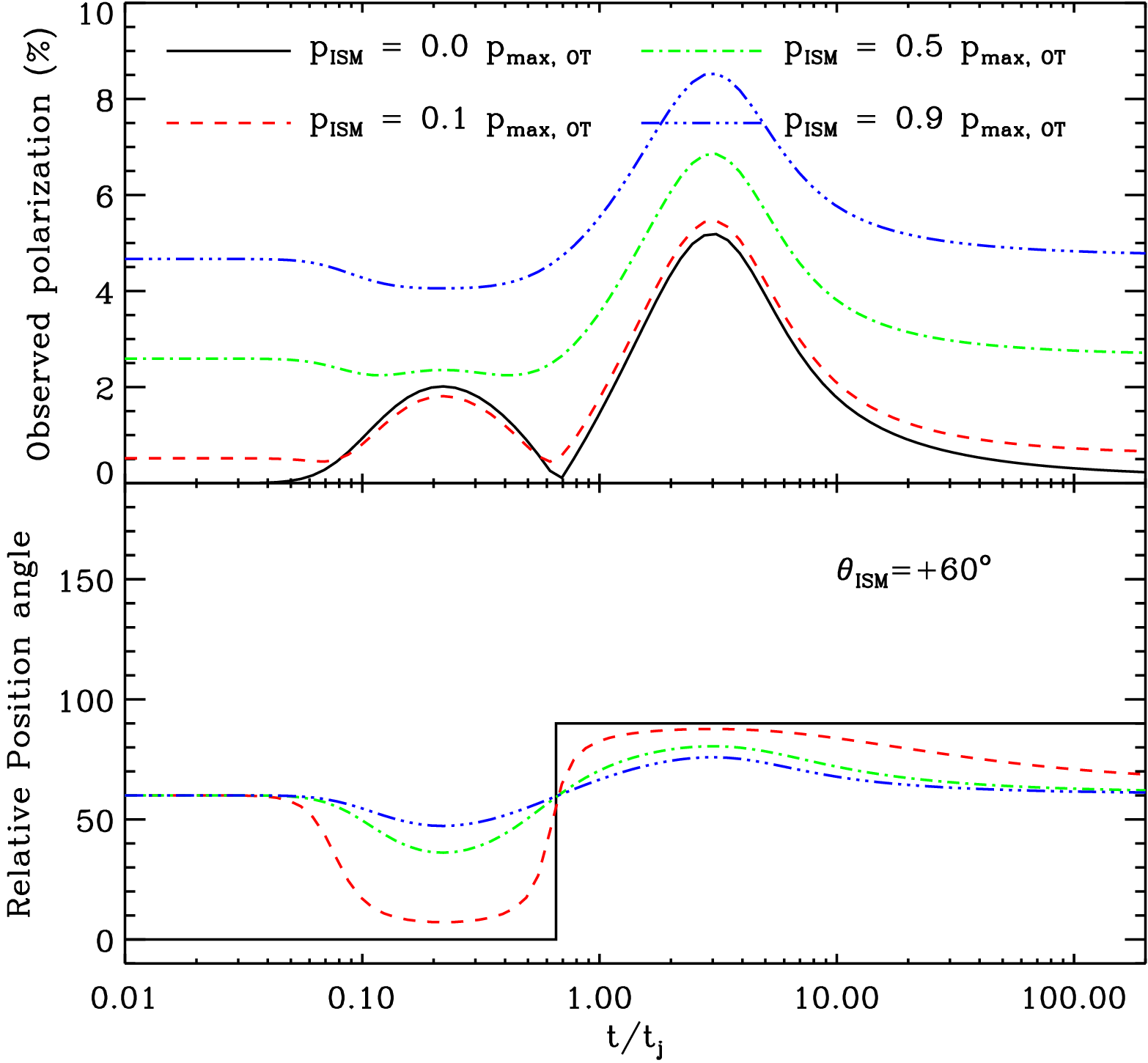}
\caption{Effect of a polarizing ISM on the theoretical polarization 
of afterglows. The theoretical curves (top: polarization level;
bottom: position angle) from GL99 for $\vartheta_\mathrm{o} /
\vartheta_\mathrm{j} = 0.5$ are shown, in both panels, as solid lines
($p_{\rm ISM} = 0$). Lines with different styles show the effect of the
selective extinction for ISM with small, intermediate and comparable
polarization with respect to the OT maximum polarization. The position
angles of the ISM are reported relative to the initial angle of the
OT. The left panel shows the effect of an ISM that induces polarization
with a position angle $\vartheta_{\rm ISM} = -19\degr$, while the right
panel shows an ISM with $\vartheta_{\rm{ISM}} = 60\degr$.
\label{fig:ismot}}
\end{figure*}

\subsection{Combined modelling}

Since the polarization levels are small and different effects seem to
contribute at comparable degree, it is mandatory to perform a combined
modelling of the imaging and spectral polarimetry, combining the
effects of the OT polarization and of the MW ISM selective
extinction. The need of at least two polarizing components is due to
the presence {\it a)} of time variability of the measured polarization
(OT component) and {\it b)} of significant polarization in the field
stars (MW ISM component).

\subsubsection{Transmission of variably polarized light in a polarizing ISM}

In order to combine the effects of ISM selective extinction with the
evolving intrinsic polarization of the OT, we adopt a Mueller calculus
approach (see e.g. di Serego Alighieri \cite{dS97} and references
therein).  In this formalism, the transmitted Stokes vector
$S^\prime\equiv(I^\prime,Q^\prime,U^\prime,V^\prime)$ is computed from
the incident one $S\equiv(I,Q,U,V)$ through a matrix, called ``Mueller
matrix", which incorporates all the properties of the transmitting
medium: $S^\prime=M \cdot S$. Since developing a complete treatment of
the polarizing properties of the ISM is far beyond the scope of this
paper, here and in the following we adopt a simplified version of the
Mueller matrix which is correct for a non-birefringent dichroic
medium. With this simplification we assume that the ISM is not able to
induce circular polarization to any incident light, either polarized
or not. In fact, the ISM does induce a small degree of circular
polarization even in unpolarized sources (Martin \& Angel
\cite{MA76}), which indicates that a non-coaxial birefringent and
dichroic medium should be considered. However, the optical properties
of such a medium are described by five parameters (extinction,
orientation of the optical axes of birefringence and of dichroism and
the two respective refraction indexes) while we can constrain only
four parameters observationally (extinction, induced linear
polarization, position angle and induced circular polarization). It is
therefore not feasible to derive a complete Mueller matrix without
going into a detailed modelling of the structure and geometry of dust
grain and their alignment.

Consider a dichroic ISM that induces a polarization
$p_{\rm{ISM}}\equiv(q^2+u^2)^{1/2}$ on unpolarized stars.  Its Mueller
matrix has then the form:
\begin{equation}
M = e^{-\tau} 
\left( \begin{array}{cccc}
  1 & q & u & 0 \\
  q & {q^2+Au^2}\over{p_{\rm{ISM}}^2} & {qu(1-A)}\over{p_{\rm{ISM}}^2}  & 0  \\
  u & {qu(1-A)}\over{p_{\rm{ISM}}^2}  & {u^2+Aq^2}\over{p_{\rm{ISM}}^2} & 0  \\
  0 & 0        & 0       & A
\end{array} \right)
\label{eq:mue}
\end{equation}
where $A \equiv \sqrt{1-p_{\rm{ISM}}^2}$, and $e^{-\tau}$ is the
opacity of the medium to non-polarized radiation. In order to conserve
energy (i.e. not to have an increased transmitted intensity), this
parameter must satisfy:
\begin{equation}
e^{-\tau} \le {{1}\over{1+|q|+|u|}},
\end{equation}
the equality holding for a perfect polarizing medium, i.e. one that does
not absorb any radiation completely polarized with its same angle.  It
can also be easily shown that if $p_{\rm{ISM}}\ll1$ and $Q,U,V\ll1$ the
matrix computation is equivalent to a simple sum of the incident and ISM
normalized Stokes parameters: $Q' \approx q + Q$, $U' \approx u +
U$. Formally, the Mueller matrix in Eq.~\ref{eq:mue} is not valid in the
case of a birefringent medium. It can however be shown that it is a good
approximation (deviations $<10\%$) if the incoming light is moderately
polarized ($P<10\%$), the linear and circular interstellar polarizations
are small ($p_{\rm{ISM}}<10\%$ and $v_{\rm{ISM}}/p_{\rm{ISM}}<10^{-2}$;
Martin \& Angel~\cite{MA76}) and the circular polarization induced on
polarized sources is small ($V/P<10^{-1}$ for active galactic nuclei;
Landstreet \& Angel~\cite{LA72}).

In order to exemplify the effect of a polarizing ISM on the intrinsic
OT polarization, we adopt the polarization model by GL99.  This model
predicts a polarization curve characterised by two distinct peaks,
whose absolute intensity depend on the angle between the line of sight
and the axis of the fireball (assumed to be collimated in a jet).  The
polarization position angle is rotated by 90 degrees at the time of
null polarization between the two peaks.  This model is illustrated by
the solid lines in Fig.~\ref{fig:ismot} (i.e. no ISM polarization).
With different line-styles, this figure also shows how the predicted
polarization degree and polarization angle change in time once
modified by some intervening polarizing ISM.  All angles are reported
relative to the initial OT polarization angle.  In the left panel the
case relative to this burst is shown ($\vartheta = -18\degr$; see
below), while in the right panel a larger misalignment between the OT
and ISM angles is shown. The figure shows that the presence of the ISM
modifies quite substantially the observed polarization, especially at
times when $P_{\rm OT} \approx P_{\rm ISM}$. Note also that the
presence of the polarizing ISM makes the position angle to vary
smoothly, instead of sharply.

\subsubsection{Modelling of the polarization curve}

To model the time-resolved polarization measurements in the framework
of available models (GL99) with a polarizing ISM we perform a fit to
the polarization curve propagating the intrinsic OT polarization
through the MW ISM with the use of the Mueller matrix derived in
Eq.~\ref{eq:mue}. The model has 6 degrees of freedom: the initial
position angle of the OT intrinsic polarization, the off-axis angle of
the line of sight, the degree of alignment of the magnetic field, the
jet break time and the ISM $q$ and $u$ parameters. The last three
parameters can be constrained with the observations, but we let them
free to vary here to check {\it a posteriori} the agreement of the
derived values with the observations. In addition, we adopt models
with and without sideways expansion of the jet (GL99; Sari~\cite{S99};
Rossi et al. 2003, in prep). A particularly important issue is to
check whether the prediction of the models are consistent with the
position angle rotation detected by Rol et al. (\cite{Rol03}). The
fit is performed on the $q$ and $u$ parameters rather than on $p$ and
$\vartheta$ due to their better statistical properties.

\begin{figure}
\includegraphics[width=\columnwidth]{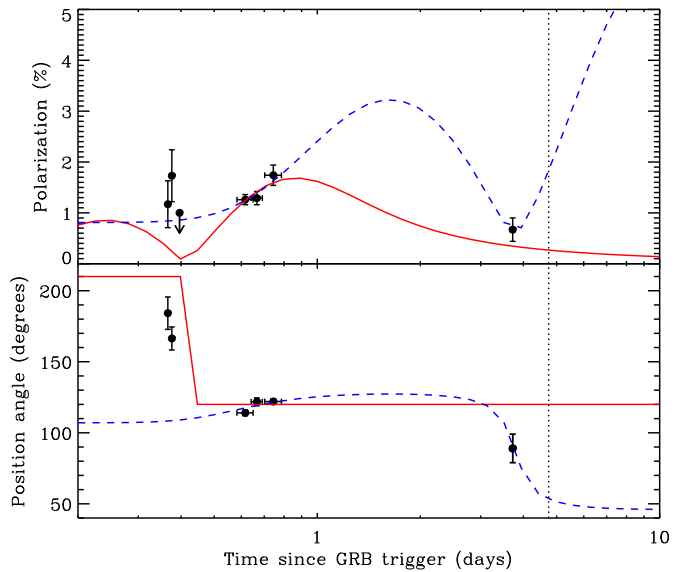}
\caption{{Time resolved modelling of the polarization and position
angle of GRB\,021004. The first three and the fifth datum are taken
from Rol et al. (\cite{Rol03}). Our NIR upper limit is not included.
The solid line show the best fit obtained from the whole dataset, with
a free break time and free properties of the ISM polarization. Even
with this extra freedom a good fit cannot be obtained due to the
rapidity of the evolution of the polarization (see text for more
details). The dashed line shows instead the best fit model obtained by
modelling only the last four data. In this case the model yields
acceptable $\chi^2/\rm{d.o.f.}=6.3/4$ and a best-fit break time
comparable to the one derived from the lightcurve fitting (see text
for more details). The vertical dotted line shows the break time
$t_b=4.74$~d derived from lightcurve modelling (Holland et
al. \cite{H03}).}
\label{fig:polafit}}
\end{figure}

The result of the fit is that it is not possible to model the
$\sim90\degr$ rotation of the position angle in the framework of the
proposed models, not even with the addition of a polarizing ISM with
free properties. In fact, the smallest $\chi^2$ that can be obtained
is $\chi^2=43$ for 8 degrees of freedom (for a non sideways expanding
jet). In addition, the fit formally yields a best break time
$t_{\rm{j}}=0.25$~d, in disagreement with the value
$t_{\rm{j}}=4.74^{+0.14}_{-0.8}$ inferred from the lightcurve
modelling (Holland et al.~\cite{H03}). Also, the best fit excludes the
presence of a sizably polarizing ISM, in disagreement with the
observation of polarization of the stars in the field of
GRB\,021004. This could be explained with a polarizing ISM in the host
galaxy with properties opposite to those of our ISM. Such a
possibility would however require an unacceptable fine tuning and is
ruled out by the modelling of the spectropolarimetric observation (see
\S~3.2). The formal best fit model, re-converted in polarization and
position angle, is shown in Fig.~\ref{fig:polafit} with a solid line.

There is one important issue, however, that makes this result non
conclusive. The models for polarization that we have used in the fit
above are derived under the assumption that the fireball and the ISM
are homogeneous. In this case, the total lightcurve should be
characterised by a smooth broken power-law decay. This is not the case
for GRB\,021004. Its lightcurve, as shown in Fig.~\ref{fig:polc} (see
also Lazzati et al.  \cite{LR02}), shows prominent bumps overlaid on
this power-law. There are several possible explanations for these
bumps: inhomogeneities in the ISM (Lazzati et al. \cite{LR02}),
inhomogeneities in the fireball (Nakar et al. \cite{N03}) or delayed
injection of energy from the central engine (Fox et
al. \cite{F03}). While in the latter hypothesis the polarization curve
should not be affected, inhomogeneities, either in the fireball or in
the ISM, can produce a polarization signal by breaking the symmetry of
the fireball emission at early times (Granot \& K\"onigl 2003). Such
polarization would have a randomly oriented position angle, depending
on the azimuthal location of the inhomogeneity with respect to the
line of sight. A similar effect is predicted in the case of
micro-lensing events (Loeb \& Perna \cite {LP98}). If we look more
closely at Fig.~\ref{fig:polc}, we note that the first three
measurement of Rol et al. (\cite{Rol03}) lie on top of a major
rebrightening, while the remaining four observations were performed in
relatively unaffected time intervals. Only the last four measurements
should therefore closely follow the theoretical model.

To check this idea, we have performed the same modelling described
above only on the final four points, fixing the ISM polarizing
properties to that of the stars in the field of GRB\,021004 in order
to limit the number of free parameters. Indeed, the last four points
can be successfully described by the model, with a break time
$t_{\mathrm j} = 3\pm1$~days, in good agreement with the
$t_{\mathrm{j}}=4.74^{+0.14}_{-0.8}$~d obtained by Holland et
al. (\cite{H03}) from the break in the lightcurve. The best fit model,
which has $\chi^2 = 6.3$ for 4 degrees of freedom, is shown with a
dashed line in Fig.~\ref{fig:polafit}. The fit requires a large degree
of alignment of the magnetic field of $75\%$ and a moderate off-axis
line of sight $\vartheta_{\rm o}/\vartheta_{\rm j} = 0.45$. The best
fit model is obtained for a non sideways expanding jet. A sideways
expanding jet cannot however be rejected, with
$\chi^2/{\rm{d.o.f.}}=7.2/4$. This ambiguity is due to the fact that,
according to the best fit model, the measured points lie in the first
peak of the polarization curve, where the sideways expansion has only
a marginal effect. The best fit model requires a misalignment of
$(-19\pm1)\degr$ between the OT initial position angle and the ISM
polarization. The effect of such a misalignment is shown in the left
panel of Fig.~\ref{fig:ismot}.

\subsection{Complications}

Additional complications can be envisaged in locally reddened
afterglows. If in fact a strong polarization is induced by dust in the
close vicinity of the burst explosion site, two effects can take
place. First, dust destruction by the burst prompt and afterglow
emission (Waxman \& Draine \cite{WD00}; Perna \& Lazzati~\cite{PL02})
can imprint a strong temporal evolution in the host ISM
polarization. This is however unlikely to be relevant for our
observations, since the typical time scale for this process is of the
order of minutes, rather than days. Second, a patchy absorber may
obscure in a different way different portions of the fireball
(e.g. Wang et al. \cite{W03}), altering one of the key assumptions of
the models. None of these effects should alter dramatically the
lightcurve, but a random noise could be overlaid on the smooth
theoretical evolution of the polarization predicted by the models.

In addition, a preferential direction for the polarization, different
from the one defined by the plane containing the jet axis and the line
of sight, can be defined by the presence of an interstellar magnetic
field of sufficient magnitude. Such a possibility has been recently
studied by Granot \& K\"onigl (\cite{GK03}). Qualitatively, the effect
of such a pre-existing field is not different from that of a
polarizing ISM. Due to the presence of a second asymmetry, the
evolution of the polarization angle is smooth rather than sharp, and
the external component dominates the observed polarization properties
when the intrinsic OT one is small. Such additional degrees of freedom
were not necessary in our last fit (nor they can explain the early
$\sim90\degr$ angle rotation). We stress however that, since all our
data were observed before the jet break, we cannot exclude the
presence of an ISM magnetic field with a well defined orientation.

\section{Summary and conclusions}

We have presented multi-time and multi-filter observations of the
polarization of the afterglow of GRB\,021004 performed with ESO--VLT
and TNG as well as our analysis of the publicly available ESO--VLT
spectropolarimetric observation.  The interpretation of the
observations is complex since none of the polarizing mechanisms that
contribute to the observed polarization seems to clearly dominate over
the others. We therefore adopt this afterglow as a case study to
investigate and describe these three main effects: intrinsic (and time
varying) OT polarization, host ISM polarization, and MW ISM
polarization. By modelling the spectropolarimetric and time-resolved
imaging polarimetry we were able to get rid of the contribution of the
host ISM, while OT and MW ISM polarizations seem to play an
intertwined role, one dominating over the other at different times.

To perform a detailed time dependent modelling of the polarization and
position angle evolutions, we implemented our data-set with the four
observations of Rol et al. (\cite{Rol03}). The complete dataset is
particularly interesting since a sizable rotation of the position
angle is present. Our attempt to fit the polarization data within the
framework of GL99 models was however a failure, given the short
timescale of the evolution and the lack of an appreciable break in the
lightcurve at the time of the position angle evolution. Since the
angle rotation is associated to one of the rebrightening events in the
lightcurve, this burst is not suited for a comparison with models,
which are computed for homogeneous fireballs producing featureless
lightcurves. A possible explanation within exsisting models is that
the early time polarization is dominated by a local enhancement of the
fireball emission, which breaks the symmetry of the fireball producing
a polarized signal. Indeed, the late time polarization data can be
successfully and consistently described by the models. We therefore
suggest that the lightcurve bumps are due to local events on the
fireball surface, such as inhomogeneities in the ISM (Lazzati et
al. \cite{LR02}) or within the fireball itself (Nakar et
al. \cite{N03}). Refreshed shocks (Fox et al. \cite{F03}) are instead
unable to account for the combined lightcurve and polarization
evolution and can be rejected.

The quality of the data, and in particular the lack of late time
measurements) does not allow us to pin down the geometry and dynamic
of the outflow, but we have shown that, with good quality
spectropolarimetry and multi-time/multi-filter imaging polarimetry, it
is in principle possible to disentangle the three effects and get a
hold on the intrinsic polarization and on the structure and dynamics
of GRB outflows. The added value of such a measurement would be the
study of the polarizing properties of dust in high redshift galaxies,
a poorly studied property of such an important component of high
redshift objects.

\begin{acknowledgements}
We thank the TNG and Paranal ESO staffs for the professional, kind and
reliable support. We thank Evert Rol for providing its results in
table form. This work was partly made using public data from the ESO
VLT Science archive. The work of Scott Barthelmy in maintaining the
GCN system is invaluable and greatly appreciated. DL acknowledges
support from the PPARC postdoctoral fellowship PPA/P/S/2001/00268.
\end{acknowledgements}

\end{document}